\documentclass[twocolumn,showpacs,preprintnumbers,amsmath,amssymb]{revtex4}

\usepackage{graphicx}
\usepackage{dcolumn}
\usepackage{bm}

\begin{document}

\preprint{wigner PRA 04 2.tex}

\title{{\bf A Modified Wigner's Inequality for Secure Quantum Key
Distribution}}

\author{S. Castelletto}
 \altaffiliation[Permanent address: ]{Istituto Elettrotecnico Nazionale G.
Ferraris, Strada delle Cacce 91, 10135 Torino (Italy)}
 \email{castelle@ien.it}
\affiliation{%
Optical Technology Division\\
National Institute of Standards and Technology, Gaithersburg,
Maryland 20899-8441
}%

\author{I. P. Degiovanni}%
 \email{degio@ien.it}
\author{M. L. Rastello}

\affiliation{%
Istituto Elettrotecnico Nazionale G. Ferraris \\
Strada delle Cacce 91-10135 Torino (Italy)
}%

\date{\today}

\begin{abstract}

In this report we discuss the insecurity with present
implementations of the Ekert protocol for quantum-key distribution
based on the Wigner Inequality. We propose a modified version of
this inequality which guarantees safe quantum-key distribution.

\end{abstract}


\pacs{03.67.Dd, 03.67.-a, 03.65.Ud}

\maketitle

Following the first proposal by Bennett and Brassard
\cite{bennet&brassard} and the later introduction of the Ekert protocol
invoking
entangled states \cite{ekert}, various systems of quantum key
distribution (QKD) have been implemented and tested by groups
around the world, turning QKD into the most advanced
application in quantum information science.

QKD offers the possibility that two remote parties, conventionally
called Alice and Bob, exchange a secret random key to implement a
secure encryption-decryption algorithm, without meeting
\cite{bennet&brassard,ekert,bb92}. QKD provides a significant
advantage over the public-key cryptography because the security of
the distributed key relies on the laws of quantum physics
\cite{bennet&brassard,ekert,bb92}, i.e. the wave packet collapse
prohibits gaining information from a quantum channel without
disturbing it. Indeed, any attempt by a third party (Eve) to
obtain information about the key is detected.

Two main goals underly the implementation of QKD schemes. One is
to create and preserve authentic quantum channels against
decoherence effects induced by any interaction with the
environment \cite{gisinrevmod}, eventually reducing or even
destroying invulnerability of quantum channels against Eve's
attack. The other is to provide a true guarantee of absolute
security against any possible eavesdropping attack i.e. the
security is not simply based on technological feasibility.

Scientists are currently
using one of two  photon sources for practical QKD, faint lasers
\cite{faint1,faint2,faint3,faint4,faint5,faint6} or  spontaneous parametric
down-conversion to generate entangled photon states
(SPDC) \cite{ekertrarity,sasha,qk2,qk3,qk1}.

Both schemes have disadvantages, although
Brassard \textit{et al}. \cite{brassard} proved theoretically that
QKD schemes based on entangled photons offer enhanced performance
and security relative to schemes based on
weak coherent pulses. In fact, entangled states allow a further
test of security based on the completeness of quantum mechanics;
in other words, the Ekert protocol exploits either the
Clauser-Horne-Shimony-Holt (CHSH) \cite{ekert} or the Wigner
inequality \cite{qk2} as an ultimate test of eavesdropping
attack.

Wigner's inequality was originally intended \cite{qk2} to provide
an easier but  as equally reliable eavesdropping check as the CHSH
test in implementing the Ekert protocol. In this report we
disclose the weakness of the Wigner inequality as a security test
when Eve has the total control of quantum channels.

\begin{figure}[tbp]
\par
\begin{center}
\includegraphics[angle=0, width=8 cm, height=5 cm]{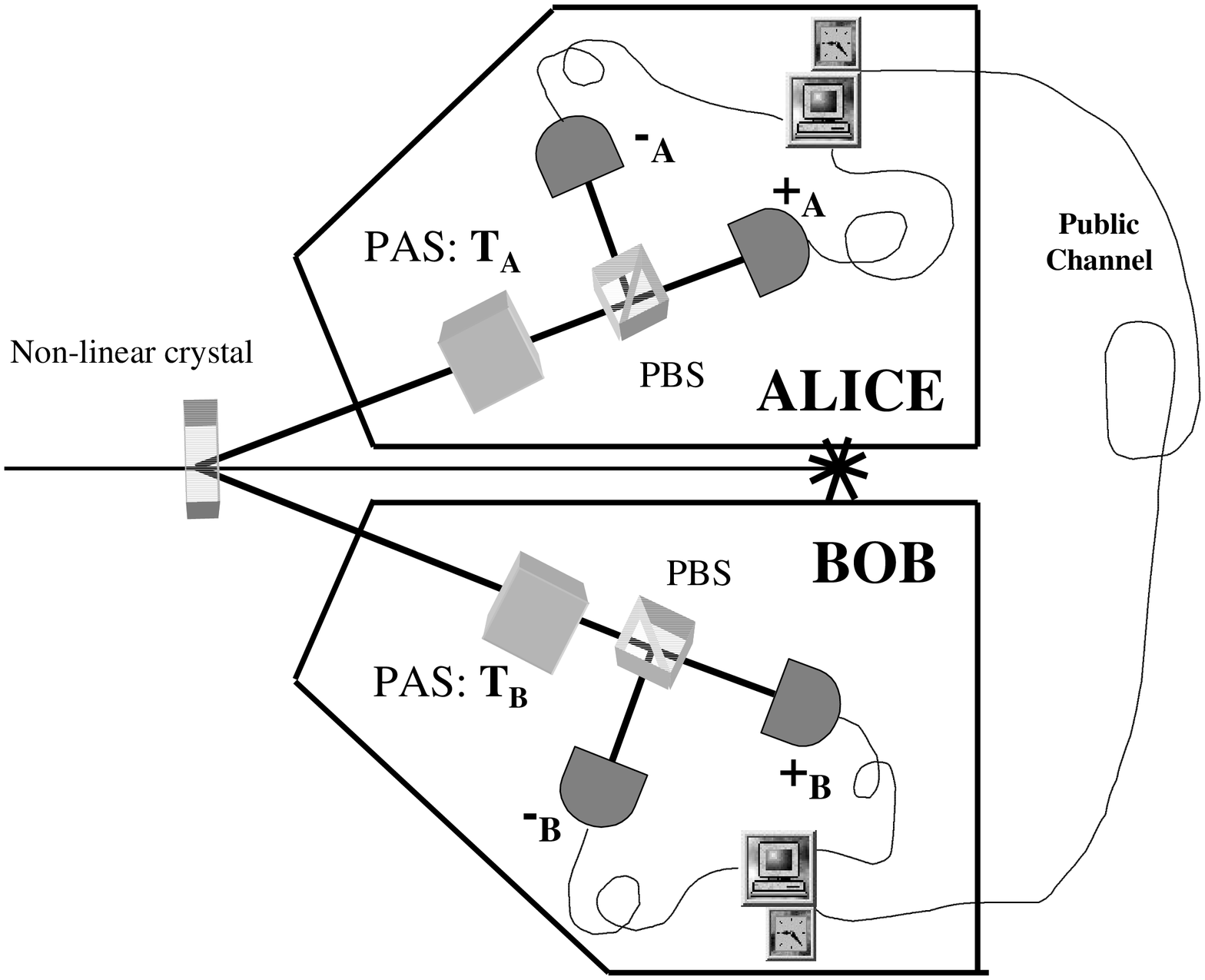}
\end{center}
\caption{ QKD set-up: polarization entangled photons generated by
SPDC\ are directed to the two parties (Alice and Bob). The bit
sequence of the key is obtained by means of polarization sensitive
synchronized measurements performed by Alice and Bob according to
a specific QKD protocol. } \label{Figure 1}
\end{figure}

In Fig. 1 we depict a typical scheme for QKD by means of entangled
photons. QKD performed by pure entangled states relies on the
realization of two quantum-correlated optical channels yielding
quasi single-photon polarization states. Alice's measurement of
the polarization of a photon of the pair, automatically projects
Bob's photon into a defined polarization state, and reversibility.
Alice's and Bob's detection apparatuses are built using
polarization analyzer systems (PAS), polarizing beam splitters
(PBS), detectors ($+_{A},\,-_{A},\,+_{B},\,-_{B}$),  data storage
systems (computers), and synchronization systems.

Consider, as an example, type II parametric down-conversion
entangled states \cite{kwiat2}, where the output two-photon states
are a quantum superposition of orthogonally polarized photons,
i.e. in the singlet state
\[
\left| \psi ^{-} \right\rangle =\frac{1}{\sqrt{2}}\left( \left|
H_{A}\right\rangle \left| V_{B}\right\rangle -\left|
V_{A}\right\rangle \,\left| H_{B}\right\rangle \right) .
\]

Without any loss of generality, a particular QKD procedure can be described
by the transformations, $%
\widehat{T}_{z}$,with $z=A,B$, on the
polarization state of a photon, e.g. a rotation of angles
$\alpha_{A}$ and $\alpha_{B}$ respectively according to
\begin{eqnarray*}
\widehat{T}_{z}\left| H_{z}\right\rangle &=&\cos \alpha_{z}\left|
H_{z}\right\rangle
+\sin \alpha_{z}\left| V_{z}\right\rangle , \\
\widehat{T}_{z}\left| V_{z}\right\rangle &=&\sin \alpha_{z}\left|
H_{z}\right\rangle -\cos \alpha_{z}\left| V_{z}\right\rangle .
\end{eqnarray*}

We assume ideal PBS in channel $z$ transmitting photons in the
state $\left| H_{z}\right\rangle $ towards detectors $+_{z}$ and
reflecting photons in the state $\left| V_{z}\right\rangle $
towards detector $-_{z}$.

The joint probabilities of each possible pair of detectors firing due to
the arrival of a photon pair with PAS angles of $\alpha_{A}$ and
$\alpha_{B}$ are
given by
\begin{eqnarray*}
p_{\alpha_{A} ,\alpha_{B} }(+_{A},+_{B}) &=&\left| \left\langle
H_{B}\right| \left\langle H_{A}\right|
\widehat{T}_{A}\widehat{T}_{B}\left| \psi
^{-}\right\rangle \right| ^{2} \\
&=&\frac{1}{2} \sin ^{2}(\alpha_{A} -\alpha_{B} ) \\
p_{\alpha_{A} ,\alpha_{B} }(+_{A},-_{B}) &=&\left| \left\langle
V_{B}\right| \left\langle H_{A}\right|
\widehat{T}_{A}\widehat{T}_{B}\left| \psi ^{-}\right\rangle
\right| ^{2}  \\
&=& \frac{1}{2} \cos ^{2}(\alpha_{A} -\alpha_{B} ) \\
p_{\alpha_{A} ,\alpha_{B} }(-_{A},+_{B}) &=&\left| \left\langle
H_{B}\right| \left\langle V_{A}\right|
\widehat{T}_{A}\widehat{T}_{B}\left| \psi ^{-}\right\rangle
\right| ^{2} \\
&=& \frac{1}{2} \cos ^{2}(\alpha_{A} -\alpha_{B} ) \\
p_{\alpha_{A} ,\alpha_{B} }(-_{A},-_{B}) &=&\left| \left\langle
V_{B}\right| \left\langle V_{A}\right|
\widehat{T}_{A}\widehat{T}_{B}\left| \psi ^{-}\right\rangle
\right| ^{2} \\
&=&\frac{1}{2} \sin ^{2}(\alpha_{A} -\alpha_{B} ).
\end{eqnarray*}

Here, we consider the variant of the Ekert protocol based on the
Wigner inequality originally proposed in ref. \cite{qk2}. For
the analyzer settings the possible combined choices by Alice and
Bob can be split into two groups: the first for key distribution and
the second testing security. We assume
Alice and Bob measure randomly among four combined analyzer
settings. Alice's possible choices are $\alpha_{A}=(\alpha_{A,1}
=-\pi /6,\alpha_{A,2}=0)$ and Bob's possible choices are
$\alpha_{B}=(\alpha_{B,2}=0,\alpha_{B,3}=\pi /6).$ The key is
obtained from the subset of measurements corresponding to parallel
 PAS settings (i.e. $\alpha_{A,i}=\alpha_{B,i}=0$)  and the firing detectors
$+_{A}-_{B}$ and $-_{A}+_{B}$.

The Wigner inequality is based on the Wigner parameter
\begin{eqnarray}
W =
p_{\alpha_{A,1},\alpha_{B,2}}(+_{A},+_{B})+p_{\alpha_{A,2},\alpha_{B,3}}(+_{A},+
_{B}) \nonumber \\
-p_{\alpha_{A,1},\alpha_{B,3}}(+_{A},+_{B}). \label{w1}
\end{eqnarray}
Note that for the maximally entangled states $W=-1/8$ while for
any local realistic theory $W\geq 0$.

To obtain the Wigner inequality $W\geq 0$ it is necessary to
summarize the Wigner argument \cite{wigner}. Two assumptions are
stipulated in the proofs of the Wigner inequality: locality and
realism. Locality means that Alice's measurement results do not
influence Bob's results, and \textit{vice versa}. Realism means
that, given any physical property its value exists independently
of its observation or measurement.

In the present case this is translated in terms of a classical
probability distribution, $\mathcal{P}(x_{1},x_{2};y_{2},y_{3})$,
where $x_{1}$ and $x_{2}$ are the hidden variables associated with
the physical property inducing Alice's measurement outcome in the
presence of the PAS rotation angle $\alpha_{A,i}$. Similarly
$y_{2}$ and $y_{3}$  correspond to the physical property inducing
Bob's outcomes when PAS angle is set as $\alpha_{B,2}$ and
$\alpha_{B,3}$. Thus, we can identify the possible values of
$x_{1,2}$ and $y_{2,3}$ with Alice and Bob's measurement outcomes,
in other words $x_{1,2}=+_{A},-_{A}$ and $y_{2,3}=+_{B},-_{B}$.
Following the Wigner approach we write,
\begin{eqnarray}
p_{\alpha _{A,1},\alpha _{B,3}}(+_{A},+_{B}) =\sum_{x_{2},y_{2}}%
\mathcal{P}(+_{A},x_{2};y_{2},+_{B}) \nonumber \\
=\mathcal{P}(+_{A},+_{A};+_{B},+_{B})+%
\mathcal{P}(+_{A},-_{A};-_{B},+_{B})  \nonumber \\
+\mathcal{P}(+_{A},+_{A};-_{B},+_{B})+\mathcal{P}(+_{A},-_{A};+_{B},+_{B}).
\label{WIG1}
\end{eqnarray}

In Wigner's original paper \cite{wigner},
$\alpha_{A,2}=\alpha_{B,2}$ and
$\mathcal{P}(x_{1},+_{A};+_{B},y_{3})=\mathcal{P}(x_{1},-_{A};-_{B},y_{3})=0
$. The assumption of perfect anticorrelation is obviously
reasonable in the test of realism and locality of a physical
theory, because it reflects the classical counterpart of a quantum
system prepared in the singlet state, i.e.
$p_{\alpha_{A,2},\alpha_{B,2}}(+_{A},+_{B})=p_{\alpha_{A,2},\alpha_{B,2}}(-_{A},
-_{B})=0$. Thus, the inequality $W\geq 0$ is obtained from Eq.
(\ref{WIG1}) by simply observing that
$\mathcal{P}(+_{A},-_{A};+_{B},+_{B})\leq \sum_{x_{2},y_{3}}\mathcal{P}%
(+_{A},x_{2};+_{B},y_{3})=p_{\alpha _{A,1},\alpha
_{B,2}}(+_{A},+_{B})$
and $\mathcal{P}(+_{A},+_{A};-_{B},+_{B})\leq \sum_{x_{1},y_{2}}\mathcal{P}%
(x_{1},+_{A};y_{2},+_{B})=p_{\alpha _{A,2},\alpha
_{B,3}}(+_{A},+_{B})$.

Ref.s \cite{qk2,xuo} suggest to be cautious in applying Wigner's
inequality to test the security of cryptography schemes, since
authors are aware that the Wigner's inequality is derived assuming
perfect anticorrelations, which are only approximately realized in
practical situations. Nevertheless a punctual quantification of
the insecurity induced by this assumption in the Ekert protocol
based on Wigner's inequality has never been pointed out.

In particular, after stating that the violation of the Wigner's
inequality ascertains the security of the quantum channels, Ref.
\cite{qk2} suggests to replace the Wigner's inequality by the
generalization of the Bell's inequality presented in Ref.
\cite{ryff} for a substantial deviation from perfect
anticorrelations. This last point has been overlooked by most
readers probably because it is not clear what a ''substantial
deviation from the perfect anticorrelations'' is.

When the eavesdropper, Eve, measures photons on either one or both
of Alice's and Bob's channels, her presence might be expected to
be revealed by a higher value of $W$ than the local realistic
theory limit, as it happens for the CHSH inequality \cite{ekert}.
Unfortunately this is not the case. Even if it can be proved that,
for Eve adopting \textit{intercept-resend} strategy and detecting
only one photon of the pair, the limit becomes $W_{\text{eve}}\geq
1/16$, this is not for eavesdropping on both channels, because in
this case there is no bound.

A proof of this last assertion, according to ref. \cite{ekert},
can be obtained by considering Eve preparing each particle of the
pairs separately, so that each individual particle has a well
defined polarization direction, possibly varying from pair to
pair.  We write the probability $P(\Phi _{A},\Phi _{B})$ of having
Alice's particle in state $\left| \Phi _{A}\right\rangle $, and
Bob's particle in state $\left| \Phi _{B}\right\rangle $, where
$\Phi _{A}$
and $%
\Phi _{B}$ are two angles of Eve's polarization preparation i.e.
$\left| \Phi _{z}\right\rangle =\cos \Phi _{z}\left|
H_{z}\right\rangle +\sin \Phi _{z}\left| V_{z}\right\rangle $. If
the density operator associated with Eve's pairs of photons is
$\widehat{\rho }_{\text{eve}}=\int P(\Phi _{A},\Phi _{B})\left|
\Phi _{A}\right\rangle \left| \Phi _{B}\right\rangle
\left\langle \Phi _{B}\right| \left\langle \Phi _{A}\right| $d$\Phi _{A}$d$%
\Phi _{B},$ then
\begin{widetext}
\begin{equation}
W_{eve}=\int P(\Phi _{A},\Phi _{B})\left[
\begin{array}{c}
\cos ^{2}(\Phi _{A}+\frac{\pi }{6})\cos ^{2}\Phi _{B}+\cos
^{2}\Phi _{A}\cos
^{2}(\Phi _{B}-\frac{\pi }{6})- \\
\cos ^{2}(\Phi _{A}+\frac{\pi }{6})\cos ^{2}(\Phi _{B}-\frac{\pi
}{6})
\end{array}
\right] \text{d}\Phi _{A}\text{d}\Phi _{B}.  \label{weve}
\end{equation}
\end{widetext}

In the case of single-channel eavesdropping with
\textit{intercept-resend} strategy, the anticorrelation between the
two photons in the Eve measurement base is preserved because $\Phi
_{B}=\Phi _{A}-$ $%
\pi /2$, and thus $W_{eve}\geq 1/16.$ In the most general case,
when Eve has the total control over the state of individual
particles, there is no physical bound, as $W_{eve}$ results in
some cases below the limit of local realistic theory, i.e.
$W_{eve}<0$ and surprisingly also below the quantum limit
$W_{eve}<-1/8$. For instance, if Eve gains total control of the
source of photons, she can send photons in proper polarization
states to avoid disclosure by a security test by Alice and Bob,
and for $P(\Phi _{A},\Phi _{B})=\delta(\Phi_{A}-0.6 \pi)
\delta(\Phi_{B}-0.4 \pi)$ the value $W_{eve}=-0.1995$ is below the
quantum limit.

In other words, the guarantee of Wigner's security test against
eavesdropping strategies is limited only to the detection of one
photon of the pairs while the CHSH security is independent from
the adopted strategy \cite{ekert}.

Therefore, we end up with a modified Wigner's parameter
$\widetilde{W}$ suitable for QKD security test by removing from
Wigner's original argument the anticorrelation assumption when
$\alpha_{A,2}=\alpha_{B,2}$. Starting again from Eq. (\ref{WIG1})
but rejecting the assumption
$p_{\alpha_{A,2},\alpha_{B,2}}(+_{A},+_{B})=p_{\alpha_{A,2},\alpha_{B,2}}(-_{A},
-_{B})=0$ and observing that
$\mathcal{P}(+_{A},+_{A};+_{B},+_{B})+\mathcal{P}(+_{A},-_{A};+_{B},+_{B})\leq
p_{\alpha_{A,1},\alpha_{B,2}}(+_{A},+_{B}) $,
$\mathcal{P}(+_{A},+_{A};-_{B},+_{B})\leq
p_{\alpha_{A,2},\alpha_{B,3}}(+_{A},+_{B})$ and
$\mathcal{P}(+_{A},-_{A};-_{B},+_{B})\leq
p_{\alpha_{A,2},\alpha_{B,2}}(-_{A},-_{B}),$ we obtain

\begin{eqnarray}
\widetilde{W}
&=&p_{\alpha_{A,1},\alpha_{B,2}}(+_{A},+_{B})+p_{\alpha_{A,2},\alpha_{B,3}}(+_{A
},+_{B}) \label{WIG2} \\&&+
p_{\alpha_{A,2},\alpha_{B,2}}(-_{A},-_{B})
-p_{\alpha_{A,1},\alpha_{B,3}}(+_{A},+_{B}) \geq 0. \nonumber
\end{eqnarray}
for any local realistic theory.

We finally observe that for the singlet state we still obtain
$\widetilde{W}=-1/8$ also in this case. This result happens because the
modified parameter $\widetilde{W}$ equals the original $W$ except
for the additional term
$p_{\alpha_{A,2},\alpha_{B,2}}(-_{A},-_{B})$, which is zero in the
case of singlet state.

To demonstrate the robustness of this security test against an
eavesdropping attack we perform a calculation analogous to Eq.
(\ref{weve})
\begin{widetext}
\[
\widetilde{W}_{eve}=\int P(\Phi _{A},\Phi _{B})\left[
\begin{array}{c}
\cos ^{2}(\Phi _{A}+\frac{\pi }{6})\cos ^{2}\Phi _{B}+\cos
^{2}\Phi _{A}\cos ^{2}(\Phi _{B}-\frac{\pi }{6}) \\
 + \sin
^{2}(\Phi _{A}) \sin ^{2}(\Phi _{B}) -  \cos ^{2}(\Phi
_{A}+\frac{\pi }{6})\cos ^{2}(\Phi _{B} -\frac{\pi }{6})
\end{array}     \label{weve2}
\right] \text{d}\Phi _{A}\text{d}\Phi _{B}. \]
\end{widetext}

Even in the general case when Eve has total control of the
polarization states of photons in the two channels, we obtain that
the minimum of $\widetilde{W}_{eve}=0.04428$, well above the limit
for local realistic theory. This result is completely equivalent
to the one obtained with the security test based on the CHSH
inequality \cite{ekert}.

Some further analysis of $\widetilde{W}$ must be considered for
the practical implementation of the Ekert protocol based on
Wigner's inequality. According to \cite{qk2}, we highlight that
the Ekert's protocol based on modified Wigner's inequality still
guarantees a simplification with respect to the one based on the
CHSH inequality, because Alice and Bob randomly choose between two
bases rather than three. Though the necessity of an experimental
evaluation of the term
$p_{\alpha_{A,2},\alpha_{B,2}}(-_{A},-_{B})$ forces Alice and Bob
to sacrifice part of the key for the sake of security. We note
that in any practical implementation of QKD protocols, Alice and
Bob distill from the noisy sifted key a nearly noise-free
corrected key by means of error correction procedures subjected to
the constraint of knowing the quantum bit error rate (QBER). Also,
the QBER is estimated at the cost of losing part of the key. Thus,
we suggest using the same sacrificed part of the key to estimate
both $\widetilde{W}$ and QBER.

\begin{table}[tbp]
\par
\begin{center}
\includegraphics[angle=0, width=6 cm, height=4 cm]{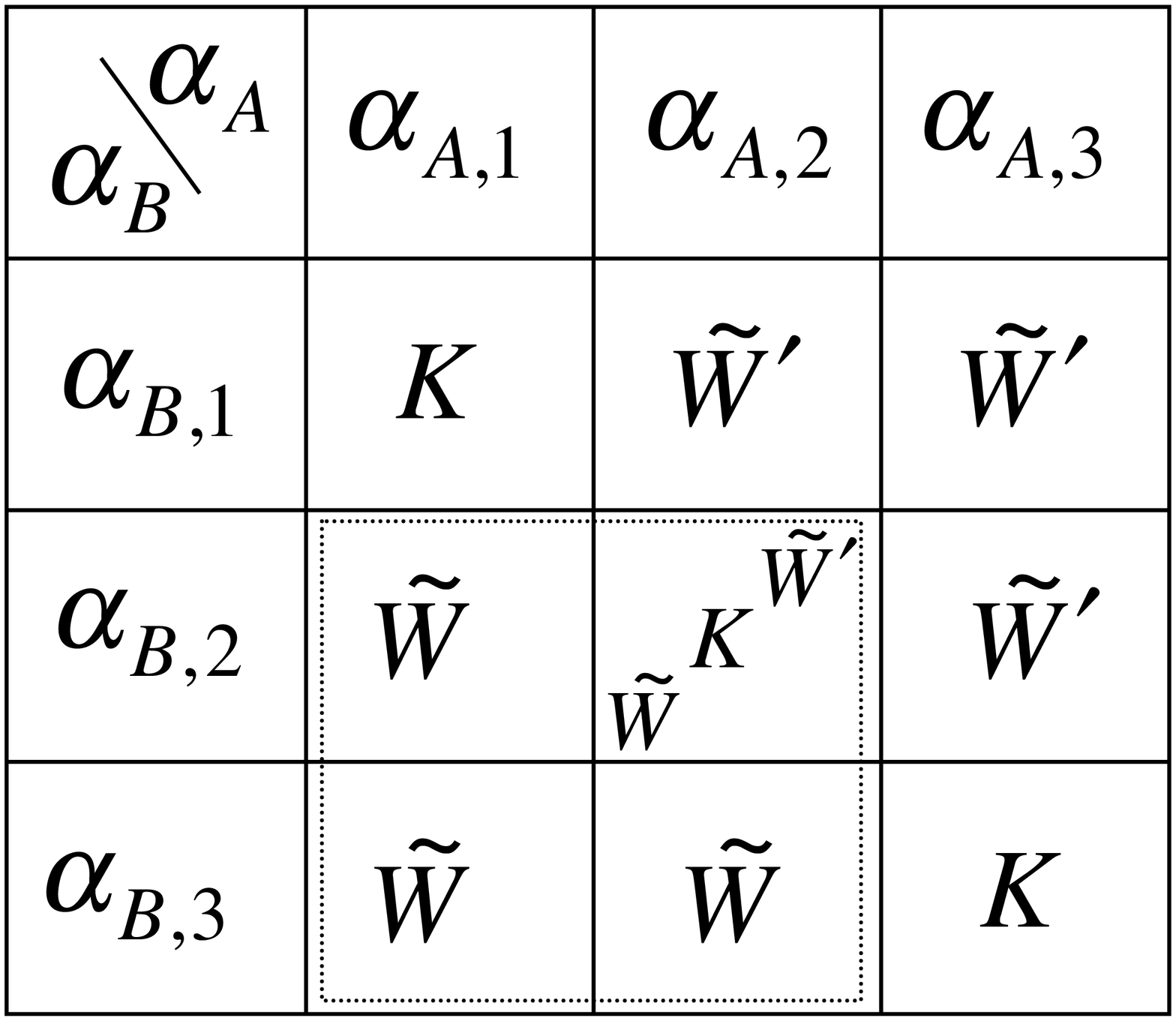}
\end{center}
\begin{caption}
{Distribution of data for Alice and Bob's analyzer settings
$\alpha_{A,1}$, $\alpha_{A,2}$ and $\alpha_{B,2}$, $\alpha_{B,3}$
respectively. The dotted square is the original protocol of T.
Jennewein, C. Simon, G. Weihs, H. Weinfurter and A. Zeilinger,
Phys. Rev Lett. {\bf 84}, 4729 (2000). \label{Table}}
\end{caption}
\end{table}

To compare the performances of the Ekert protocol based on
Wigner's inequality versus the one CHSH-based \cite{ekert} we
consider a protocol (Table I) where Alice and Bob measure randomly
using three analyzer settings (as in the case of CHSH), where
$\alpha_{A,1}=\alpha_{B,1}=-\pi/6$, $\alpha_{A,2}=\alpha_{B,2}=0$
and $\alpha_{A,3}=\alpha_{B,3}=\pi/6$. $\widetilde{W}$ and
$\widetilde{W}'$ correspond to two distinct test parameters and
$K$ to the key distribution. This protocol is more efficient than
the protocol based on CHSH. In terms of key generation we observe
that in the case of CHSH only 2/9 of the qubits exchanged are
devoted to the key generation \cite{ekert} while in the case of
our protocol this quantity is a number between 2/9 and 1/3,
depending on the security needs.

Even if, for some strict security request, all the qubits
exchanged by the two parties with analyzer settings
$\alpha_{A,2}$, $\alpha_{B,2}$ are devoted to the evaluation of
$\widetilde{W}$ ($\widetilde{W}'$), and also for this protocol
only 2/9 of the data still contribute to the key, in this protocol
none of the qubits exchanged are discarded (Table I) while in the
case of CHSH 1/3 of the qubits are discarded.

The results so far obtained allow us to quantify the so called
''substantial deviation from prefect anticorrelations'' as that
value of the QBER such that the quantum key distribution fails,
because the key is insecure even if the Wigner's inequality is
violated.

For the original protocol introduced in \cite{qk2} the QBER of the
key is the probability of correlated results when analyzer
settings are $\alpha_{A,2},\alpha_{B,2}$, i.e.
\begin{equation*}
\mathrm{QBER} =p_{\alpha_{A,2},\alpha_{B,2}}(-_{A},-_{B}) +
p_{\alpha_{A,2},\alpha_{B,2}}(+_{A},+_{B}).
\end{equation*}

Thus, this leads to
\begin{equation*}
0\leq \widetilde{W}\leq
W+p_{\alpha_{A,2},\alpha_{B,2}}(-_{A},-_{B}) +
p_{\alpha_{A,2},\alpha_{B,2}}(-_{A},-_{B}),
\end{equation*}
which can be easily rewritten in terms of QBER as
\begin{equation}
\mathrm{QBER}+W \geq 0.    \label{qpw}
\end{equation}
Thanks to the modified Wigner's inequality a limit is found for
both $W$ and the anticorrelation check. Thus, for the Ekert's
protocol based on the original Wigner's inequality a secure key
distribution is provided  when, instead of the violation of the
Wigner's inequality ($ W \geq 0$), the Wigner's parameter $W$
satisfies $W \leq -$QBER. Eq. (\ref{qpw}) and the related
considerations obviously still hold for the enlarged protocol
presented in Table I. Here one has to consider only the QBER of
that part of the key obtained with analyzers settings
$\alpha_{A,2},\alpha_{B,2}$, and not the QBER of the whole key.

As a final remark we highlight that our approach not only solves
the problem of estimating the ''substantial deviation from the
perfect anticorrelations'' but completely overtakes it. Following
our approach the experimentalist has simply to check the modified
Wigner's inequality from his experimental data to guarantee
non-locality, i.e. security of the quantum channels as in the
original Ekert's idea \cite{ekert}, without any further
anticorrelation check.

Furthermore, the experimentalist in a noisy environment can
discard a secure key when Eq. (\ref{qpw}) is used instead of the
modified Wigner's inequality, since $\widetilde{W}\leq W+$QBER.

In conclusion, this paper discusses the security of Ekert's
protocol based on the Wigner inequality. We showed that the QKD
Ekert protocol based on Wigner's inequality presents a serious
lack of security against some eavesdropping strategies other than
\textit{intercept-resend}. We emphasized the motivation beneath
the missing security and propose a modified test ultimately
guaranteeing secure QKD.

We would like to thank M. Rasetti and I. Ruo Berchera, G. Di
Giuseppe, A. M Colla, F. Bovino, P. Varisco for useful discussion
and helpful suggestions. This work was developed in collaboration
with Elsag S.p.A., Genova (Italy), within a project entitled
"Quantum Cryptographic Key Distribution" co-funded by the Italian
Ministry of Education, University and Research (MIUR) - grant n.
67679/ L. 488. In addition S. C. acknowledges the partial support
of the DARPA QuIST program and M. L. R. acknowledges the partial
support by INFM.


\begin{thebibliography}{99}
\bibitem{bennet&brassard}  C. Bennett and G. Brassard, in {\it{Proceedings of
the IEEE International Conference on Computers, Systems and Signal
Processing, Bangalore}}, (IEEE, New York 1984), p. 175.
\bibitem{ekert}  A. K. Ekert, Phys. Rev. Lett. {\bf 67}, 661 (1991).
\bibitem{bb92}  C. H. Bennett, G. Brassard and N. D. Mermin, Phys. Rev.
Lett. {\bf 68}, 557
(1992).
\bibitem{gisinrevmod} N. Gisin, G. Ribordy, W. Tittel and H. Zbinden, Rev.
Mod. Phys. {\bf 74}, 145 (2002).
\bibitem{faint1} C. H. Bennett, F. Bessette, G. Brassard, and  L.
Salvail, J. Cryptology {\bf5}, 3 (1992).
\bibitem{faint2} A. Muller, J. Breguet, and N. Gisin, Europhysics
Lett. {\bf 23}, 383 (1993).
\bibitem{faint3} P. Townsend, J. G. Rarity, and P. Tapster, Electron. Lett.
{\bf
29}, 1291 (1993)
\bibitem{faint4} A. Muller, H. Zbinden, and N. Gisin, Nature
{\bf 378}, 449 (1995).
\bibitem{faint5} B. Jacobs, J. Franson, Opt. Lett. {\bf 21},
1854 (1996).
\bibitem{faint6} W. T. Buttler \textit{et al.}, Phys. Rev. Lett. {\bf 81},
3283 (1998).
\bibitem{ekertrarity}  A. K. Ekert, J. G. Rarity, P. R. Tapster and G. M.
Palma, Phys. Rev. Lett. {\bf 69}, 1293 (1992).
\bibitem{sasha}  A. V. Sergienko, M. Atature, Z. Walton, G. Jaeger, B. E. A.
Saleh and M. C. Teich, Phys. Rev. A {\bf 60}, 2622 (1999).
\bibitem{qk2}  T. Jennewein, C. Simon, G. Weihs, H. Weinfurter and A.
Zeilinger, Phys. Rev Lett. {\bf 84}, 4729 (2000).
\bibitem{qk3}  D. S. Naik, C. G. Peterson, A. G. White, A. J. Berglund and
P. G. Kwiat, Phys. Rev. Lett. {\bf 84}, 4733
(2000).
\bibitem{qk1}  W. Tittel, J. Brendel, H. Zbinden and N. Gisin, Phys.
Rev. Lett. {\bf 84}, 4737 (2000).
\bibitem{brassard}  G. Brassard, N. Lutkenhaus, T. Mor and B. C. Sanders,
Phys. Rev. Lett. {\bf 85}, 1330 (2000).
\bibitem{kwiat2}  P. G. Kwiat, K. Mattle, H. Weinfurter, A. Zeilinger, A. V.
Sergienko and Y. Shih, Phys. Rev. Lett. {\bf 75}, 4337 (1995).
\bibitem{wigner} E. P. Wigner, Am. J. Phys {\bf 38}, 1005 (1970).
\bibitem{xuo} P. Xue, C. F. Li, and G. C. Guo, Phys. Rev. A {\bf 65}, 034302
(2002).
\bibitem{ryff} L. C. Ryff, Am. J. Phys {\bf 65}, 1197 (1997).


\end{thebibliography}
\end{document}